\documentclass[10pt,journal,compsoc]{IEEEtran}
\usepackage{lineno,hyperref}
\usepackage[table]{xcolor}
\usepackage{adjustbox}
\modulolinenumbers[5]
\usepackage{booktabs}
\usepackage{array,makecell}
\setcellgapes{1pt}
\makegapedcells
\newcolumntype{R}[1]{>{\raggedleft\arraybackslash }b{#1}}
\newcolumntype{L}[1]{>{\raggedright\arraybackslash }b{#1}}
\newcolumntype{C}[1]{>{\centering\arraybackslash }b{#1}}
\usepackage{gensymb}
\usepackage{color}
\usepackage{tikz}
\usetikzlibrary{arrows.meta,
	positioning,
	quotes}
\usetikzlibrary{matrix,shapes,arrows,positioning,chains,calc}
\usetikzlibrary{matrix,decorations.pathreplacing, calc, positioning,fit}
\usetikzlibrary{shapes.geometric,arrows}
\usetikzlibrary{shapes.geometric,arrows}
\date{}
\usepackage[all,error]{onlyamsmath}
\usepackage{amsmath,amssymb,amsfonts}

\usepackage{graphicx} 
\usepackage{gnuplottex}
\usepackage{color}
\usepackage{enumitem}
\usepackage{authblk}
\usepackage{xcolor}
\usepackage{soul}
\usepackage{amssymb}
\usepackage{pifont}
\newcommand{\cmark}{\ding{51}}%
\newcommand{\xmark}{\ding{55}}%
\usepackage{lineno}
\usepackage{array}
\ifCLASSOPTIONcompsoc
  \usepackage[nocompress]{cite}
\else
  \usepackage{cite}
\fi
\ifCLASSINFOpdf

\else

\fi

\begin{document}
\title{Event Detection for Non-intrusive Load Monitoring using Tukey's Fences}

\author{Sidi Mohammed Kaddour ,
        Mohamed Lehsaini,
        Abdelhamid Bouchachia,%
}

\markboth{Draft paper}%
{Kaddour et al.: Event Detection for Non-intrusive Load Monitoring using Tukey's Fences}

\IEEEtitleabstractindextext{%
\begin{abstract}
The primary objective of non-intrusive load monitoring (NILM) techniques is to monitor and track power consumption within residential buildings. This is achieved by approximating the consumption of each individual appliance from the aggregate energy measurements. Event-based NILM solutions are generally more accurate than other methods. Our paper introduces a novel event detection algorithm called Tukey's Fences-based event detector (TFED). This algorithm uses the fast Fourier transform in conjunction with the Tukey fences rule to detect variations in the aggregated current signal. The primary benefit of TFED is its superior ability to accurately pinpoint the start times of events, as demonstrated through simulations. Our findings reveal that the proposed algorithm boasts an impressive 99\% accuracy rate, surpassing the accuracy of other recent works in the literature such as the Cepstrum and $\chi ^2$ GOF statistic-based analyses, which only achieved 98\% accuracy.
\end{abstract}

}

\maketitle

\section{Introduction}
	
The world is facing a growing sense of urgency surrounding energy conservation efforts. It has been estimated that household energy usage accounts for approximately 33\% of the world's total energy consumption, and raising awareness of energy consumption through feedback on electricity usage from Home Electrical Appliances (HEAs) has led to a 20\% reduction in energy usage \cite{armel2013a,shaikh2014a}. Therefore, keeping a close watch on domestic energy consumption is imperative. However, the conventional approach of using sensors on HEAs to collect data, known as Intrusive Load Monitoring (ILM), can be challenging in terms of installation and upkeep.
Conversely, Non-Intrusive Load Monitoring (NILM) has emerged as a promising methodology for producing a detailed report on household energy consumption. The appliance consumption can be identified by analyzing the primary power input loads (voltage and current). The concept of NILM was originally introduced by Hart~\cite{hart1992a}, and only one meter connected to the primary electrical panel of the house is required. The various stakeholders in the smart grid ecosystem, such as power distributors, suppliers, and consumers, receive statistical and usage data. Consumers react to this information to reduce their energy consumption, and the Home Energy Monitoring System (HEMS) can employ this data to monitor energy usage within the home. Furthermore, power producers in smart grid systems can leverage this data to create superior predictive models.
	
Due to the increased interest in the field, researchers have been experimenting with various approaches and algorithms for solving energy decomposition problems that involve neural networks, big data, soft computing, and statistical methods. In energy modeling, there has been a surge in the use of Hidden Markov Models (HMMs) with probabilistic methods gaining popularity \cite{Kolter11, parson2012, zhong2015, zhong2014}. HMMs are unsupervised models but are often trained in a supervised manner in important applications \cite{Tamposis2018}. While supervised methods are the most popular solutions \cite{Bonfigli2015}, some researchers have proposed solutions using unsupervised \cite{liu2021}, reinforcement \cite{Zaouali2022, Li_2020}, and transfer learning approaches \cite{DIncecco2019}. Supervised techniques rely on classification algorithms that require sufficient labeled data for training. In addition, approaches that are not supervised tackle the issue of obtaining labeled data, which can be challenging. Based on the frequency of sampling, NILM (Non-Intrusive Load Monitoring) techniques can be divided into two main categories as noted by Faustine (2017): low sampling frequency (LSF) approaches, which maintain a sampling rate of 1 KHz or lower, and high sampling frequency (HSF) approaches, which reach from kHz to MHz.
Datasets that have a low sampling frequency are more abundant than those with a higher frequency, as noted in \cite{Henriet2018}. This makes the former more easily accessible and prevalent. In this domain, there is also a division between event-based and non-event-based NILM, which depends on whether they utilize the recognition and categorization of transition signals or not \cite{himeur2020a, Jorde2020, Baranski2004}.

The pipeline that follows event-based methodologies incorporates four key stages: event detection, feature extraction, appliance clustering, and energy usage estimation. Event detection is the initial stage of the process, and it is, therefore, crucial for the progression of the NILM pipeline. The input features utilized in event-based approaches differ from one study to another. While some works use real and reactive power as inputs for event detection, as seen in most studies \cite{Sadeghianpourhamami2017}, others opt for alternative inputs such as current signals \cite{anderson2012a} or voltage distortion \cite{nait2017a}. An event detector serves as an abnormality recognizer within a signal capture, which aids in the analysis and modeling of physical phenomena. The literature contains several approaches to event detection in NILM, with most of these methods categorized as either supervised \cite{li2012} or unsupervised \cite{nait2017a, meziane2016}.
When it comes to accuracy, event-based solutions for non-intrusive load monitoring (NILM) have proven to be more effective than non-event-based solutions. This is primarily because certain high-efficiency appliance (HEA) power usage is minimal, and the duration of the ON state is so brief that it goes unnoticed in steady-state NILM classifiers.
	
Our aim in this paper is to present a new unsupervised event detector for Non-Intrusive Load Monitoring (NILM) called TFED, which utilizes Tukey's Fences-based outlier detection method and a fast Fourier transform (FFT) sliding window. To accommodate various load fluctuations, the threshold is adjusted dynamically. Additionally, TFED establishes representative load signatures founded on the results of event detection, which are beneficial for NILM and other applications including device malfunction detection and end-user security.

The main contributions of this paper are as follows:
\begin{itemize}
\item The proposed TFED mechanism can cope efficiently with unpredictable and complicated load changes, including high fluctuation,  using an unsupervised approach where no probabilistic model or a training process is required. 

\item The proposed TFED method combines Tukey's fences outlier detection method and fast Fourier transform to find an adaptive threshold on every current window to deal with unpredictable and complicated load changes, including high fluctuation.

\item Conducting extensive case studies on BLUED (Building-level fully labeled electricity desegregation) dataset \cite{load2015a} to demonstrate TFED's accuracy and robustness. 
\end{itemize}
The rest of the paper is organized as follows: Section ~\ref{Background} introduces background terms related to TFED, such as NILM and BLUED datasets, as well as some terms used in the literature, such as chi-square goodness of fit$\chi^2 $ GOF, Cepstrum Analysis presents related work on event detection and payload signature. The ~\ref{Contribution} section describes the proposed TEFD. The ~\ref{Simulations} section introduces and discusses the simulations using the BLUED dataset. Finally, the ~\ref{Conclusion} section summarizes this paper and gives some perspectives on future work.

\section{Background} \label{Background}

\begin{figure}[ht!]
		\centering
		\includegraphics[scale=0.6]{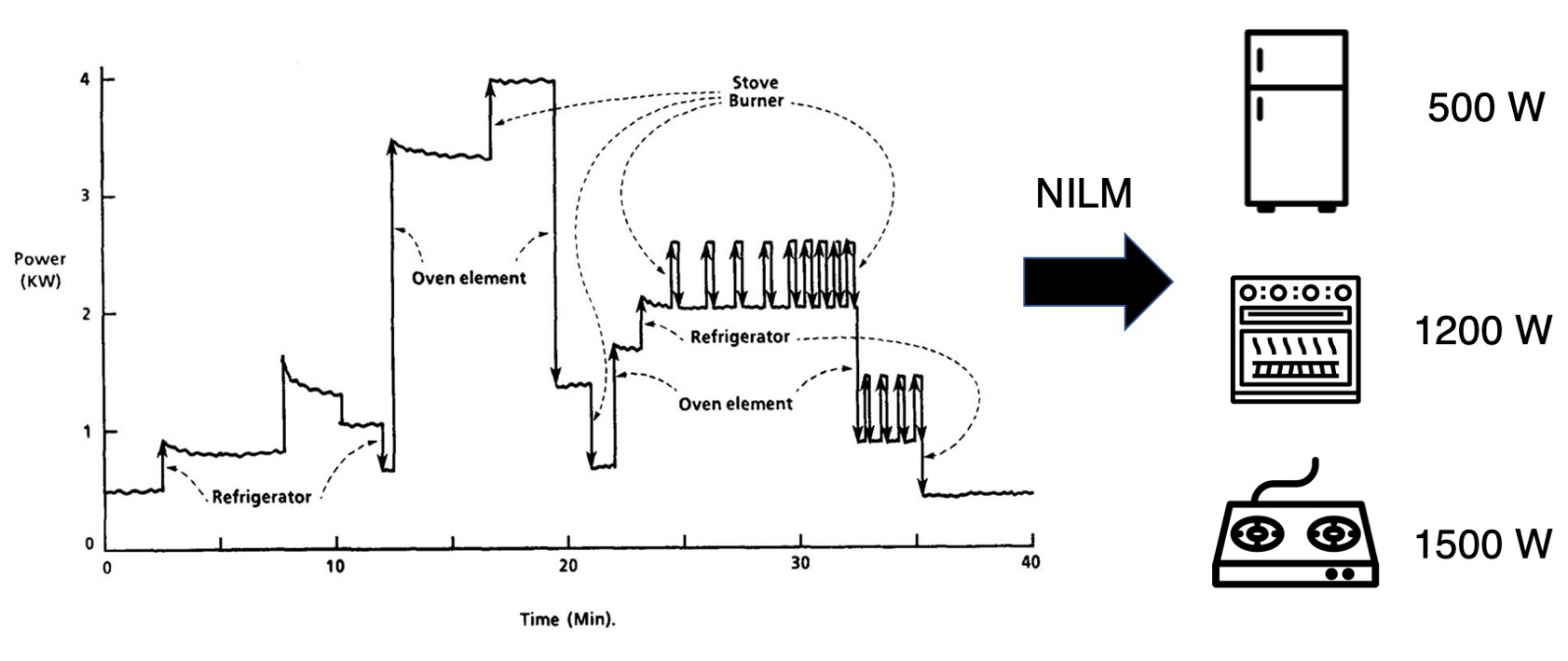}
		\caption{Power vs. time plot shows step changes due to individual appliance events \cite{hart1992a}}
		\label{fig:fig0}
	\end{figure}

Many algorithms were introduced in the context of real-time event detection for NILM \cite{DeBaets2016}. In the following, we define a few known aspects such as NILM, depicted in Fig.~\ref{fig:fig0}, and techniques commonly used for real-time event detection.

\subsection{Non-intrusive Load Monitoring (NILM)}

Non-Intrusive Load Monitoring (NILM) goal is to determine the condition of individual appliance loads, such as $P_1(t), P_2(t),..$, and so on, using the provided data of the aggregated load $P_{\mathrm{agg}}(t)$. as illustrated in the following equation.
\begin{equation}
P_{\mathrm{agg}}(t)=\sum_{i=1}^n P_i(t)
\end{equation}
The concept of NILM was initially introduced by Hart~\cite{hart1992a}. Ever since the arrival and proliferation of smart meters and their related applications \cite{Bergs2010,liang2010}, there has been a substantial increase in the number of research projects conducted on NILM. Depending on the characteristics of the appliance used for disaggregation, NILM techniques can be classified as either steady-state or transient features. The following classification level involves Learning and Inference methods, which can be either supervised, unsupervised, or both.
\subsection{BlUED dataset}

Numerous datasets have been furnished for the purpose of evaluating energy consumption and identifying the individual appliances that contribute to it in residential buildings. The datasets are divided into categories based on their sampling frequency, with datasets that have a frequency in Hz being low-frequency and primarily used for energy disaggregation. High-frequency datasets, on the other hand, have a sampling frequency of kHz and above. Choosing the correct dataset is crucial to the success of the disaggregation algorithms, and it is important to have access to both aggregated and sub-meter appliance-level datasets. The datasets examined in this study provide specific details on both aggregate and appliance-level recordings \cite{iqbal2020a}.
It is imperative to utilize a dataset with high frequency when conducting event detection. The BLUED database \cite{anderson2012b} has been selected for this purpose due to its inclusion of raw current and voltage data at a high frequency of 12 kHz for the entire house, as well as computed active power (60 Hz). The dataset also contains a list of timestamps for events that occur when the power consumption status of an appliance changes by more than 30 watts and persists for at least 5 seconds. The dataset contains data on approximately 50 appliances, and a total of 2335 events were recorded. Furthermore, the dataset features an additional 2482 events from unknown sources.

\section{Related Literature} \label{Related}
\begin{table*}[]
\centering
\begin{tabular}{@{}llcll@{}}
\toprule
Approach                 & Sampling Rate  & \multicolumn{1}{l}{Unsupervised}     & Dataset        & Data Nature\\ 
\midrule
TFED (This paper)        & High(6khz)           & \cmark         & BLUED          & C   \\

\cite{yang2014a}(GOF)             & Low(3s)             & \cmark         & REDD           & AP   \\
\cite{Wild2015}(KFDA)              & High(12khz)         & \cmark         & BLUED          & C        \\
\cite{Basu2015a}(KNN)            & Low(0.13)           & \xmark         & PRIVATE        & AP   \\
\cite{nait2017a,meziane2016}(HAND) & High(12khz)         & \cmark         & MODELED SIGNAL & C        \\
\cite{DeBaets2016} ($\chi^2$ GOF)           & Low(60hz)           & \cmark         & BLUED          & AP   \\
\cite{zhu2018a} (CUSUM)             & High(10Khz)         & \cmark         & PRIVATE        & AP   \\
\cite{lu2020a}   (GLR+LLD-MAX+AWB)           & Low(20hz)           & \cmark         & BLUED          & AP   \\
\cite{yan2021}  (WAMMA)             & Low(20-50hz)        & \cmark         & LIFTED, BLUED  & AP   \\ 
\cite{Luo2002}  (GLR)             & Low(25hz)        & \cmark         & ASHRAE  & AP   \\ 
\bottomrule
(AP): Active Power,(C) : Current
\end{tabular}
\caption{SUMMARY OF EVENT DETECTION METHODS}
\end{table*}
Approximately thirty years ago, Hart \cite{hart1992a} introduced a technique to break down electrical loads by solely examining device-specific step-wise alterations in total power consumption. Additionally, various event detection methods have been put forth. These techniques are categorized based on their type of learning, sampling frequency, and data nature. In \cite{Luo2002}, a group of researchers proposed the utilization of the Generalized Likelihood Ratio (GLR) method to calculate a decision statistic based on the logarithm of the ratio of probability distribution before and after a possible shift in the mean. Their process necessitates the offline training of four parameters, which include the duration of the moving window, the variance of performance data, and the threshold for detection statistics. This approach has been adopted and improved upon by numerous studies. According to \cite{lu2020a}, a hybrid solution was proposed by researchers, which combines a time-limited algorithm based on moving average variation with two auxiliary algorithms based on derivative analysis and filtering analysis. This solution aims to detect false events through the use of active power features with a low sampling rate and fixed parameters. However, the effectiveness of this approach may be limited in detecting time events that experience rapid transitions. To address this limitation, \cite{Wild2015} presented an unsupervised solution that utilized a sliding-window kernel Fisher discriminant analysis (KDFA) on the BLUED dataset. In this approach, an event is defined as an active session that deviates from one steady-state section to another. Since event sessions can vary in duration, accurately detecting the start and end of each event is crucial for the NILM detector. The researchers in \cite{nait2017a,meziane2016} have developed a highly precise NILM (Non-Intrusive Load Monitoring) detector named "HAND" using an unsupervised event-based algorithm. The HAND algorithm outperforms the KFDA (Kernel Fisher Discriminant Analysis) approach by achieving a detection probability of 96.7\% on simulated data. The algorithm focuses on the Envelope function of the current signal and the standard deviation, which makes it both efficient and fast. However, the simulation was evaluated using a modeled signal, where the noise parameter can be changed, which is not the case with real data readings. The authors of \cite{yang2014a,baets2017a} adopted the GOF (Goodness of Fit) approach to test whether two consecutive time frames share a typical distribution by deriving a decision function from the log-probability distribution ratio before and after a potential change in the mean value. In \cite{nguyen2014a, zhu2018a}, the authors applied a CUSUM (Cumulative Sum) algorithm to identify the beginning and end of a HEA (High-End Audio) transient active power signal. Zhu and his team \cite{Basu2015a} achieved a detection probability of 90\% using their methodology on actual data, which included 200 events from eight different types of HEAs that were both turned on and off. Yan and his colleagues \cite{yan2021} suggested a method based on dynamic time windows for detecting events. This method adjusts the size of time windows and other parameters dynamically to cope with variations in load. It also extracts characteristic load signatures based on the results of event detection, which are beneficial for NILM and other applications. Nevertheless, a low sampling rate can lead to the loss of minor trace changes. To find the corresponding appliances, some proposed solutions, such as that presented by Leslie et al. \cite{Norford1996}, utilize matched filters to correlate transient signals from known appliances with aggregated consumption signals. This idea was further advanced in \cite{DeBaets2016} by utilizing Cepstrum analysis on the power signal. However, small appliances with low fluctuations can badly affect the matched filter's performance. another related work in \cite{Rehman2020} introduces two novel algorithms specifically designed for computationally fast and low-complexity event detection. The algorithms leverage the concept of sliding windows (SW) to track statistical features of aggregated load data. To evaluate their performance, real-world data is utilized, and a comparative analysis is conducted against a recently proposed event detection algorithm. The simulations and sensitivity analysis demonstrate promising results, with the proposed algorithms achieving recall and precision rates of up to 93\% and 88\%, respectively. These findings emphasize the effectiveness of the algorithms in accurately detecting load variations and determining the corresponding events.Another method is WAMMA (Window with Adaptive Margins, Multi-window Screening, and Adaptive Threshold) proposed by \cite{Yan2023}. This particular method adjusts its parameters according to the data, which helps it maintain a high level of accuracy while remaining robust. WAMMA has been shown to outperform methods that use fixed parameters and is capable of transferring parameters to households where the data is unknown. It is also capable of capturing complete transitions, which allows for precise load disaggregation. Specific work in \cite{Kotsilitis2023} presents an algorithm that is intended to facilitate the implementation of software or hardware on-site. The algorithm's lightweight nature is the reason for its ease of use. It focuses on high-frequency sampled data and extracts simple, easily computed features while circumventing complicated operations.In this work, there are significant contributions that utilize multiple straightforward criteria in order to declare events, resulting in high detection accuracy and low computational expense. The algorithm also relies on the slope coefficient that is determined from the aggregated power data points to identify the edges of events and determine when transients end. This coefficient is a simple feature that can be easily computed, providing a trustworthy indicator of the trend in aggregated power waveforms. In addition, the algorithm uses small windows for data points, making it possible to detect events with very slight time differences, even as little as half a second, which is fitting for residential settings.\cite{Zheng2018} has addressed the use of supervised event-based approaches with harmonic current features. The additive property of these features, which ensures that appliance features remain unchanged regardless of power network states, has been acknowledged. However, little research has focused on proving this property. Event detection algorithms for NILM systems are still under investigation, and the performance of harmonic features-based NILM systems on real-life datasets is yet to be thoroughly explored. This work aims to fill these gaps by verifying the additive property of harmonic features, identifying the best event detection algorithm, and demonstrating accurate appliance identification using harmonic features-based NILM.
	
\section{Tukey's fences for event detection} \label{Contribution}

One solution for non-intrusive load monitoring (NILM) is to leverage the concept of an "event" that coincides with a change in the current signal. The process of event detection based NILM, as shown in Fig.~\ref{fig:fig1}, typically involves four stages: measuring the energy consumption (measurement), detecting events (event detection), classifying events (event classification), and estimating power usage (power usage estimation). By measuring only the primary power input in a building from a single sensing point, the system is designed to recognize alterations in the collected energy consumption data that signal a change in an appliance's state (event detection). Then, the detected transition signals can be classified (classification) to deduce the electrical energy consumption of each appliance (energy usage estimation).
	\begin{center}
		
		\tikzstyle{block} = [rectangle, draw=black, text width=90 pt, text centered, rounded corners, minimum height=25 pt]
		
		\tikzstyle{line} = [draw, -latex']
		
		\begin{tikzpicture}[node distance = 2cm, auto]
		
		\node [block] (Meseaure) {\small{Measurement collection}};
		\node[right=20 pt] [block,blue] (Dectection)  at (Meseaure.east) {\small{Event detection}};
		\node[below=50pt] [block] (Classification)  at (Dectection.south) {\small{Event classification}};
		\node[left=20pt] [block] (Energy)  at (Classification.west) {\small{Energy usage estimation}};
		
		\draw[->] (Meseaure.east) -- (Dectection.west);
		\draw[->] (Dectection.south) -- (Classification.north);
		\draw[->] (Classification.west) -- (Energy.east);
		\end{tikzpicture}
		\begin{figure}[ht!]
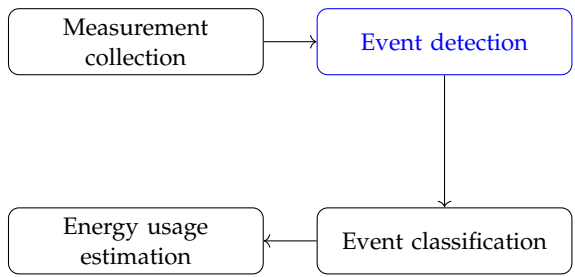

			\centering
			\caption{Event-based NILM.}
			\label{fig:fig1}
		\end{figure}
	\end{center}
Our focus is on the proposed TFED algorithm, which we will describe below. Before proceeding, however, it is imperative to establish a ground truth for the data, as this will be the basis for evaluating the algorithm's performance in detecting events during the test. The ground truth comprises a series of timestamps that pinpoint the exact moments at which events take place.
	
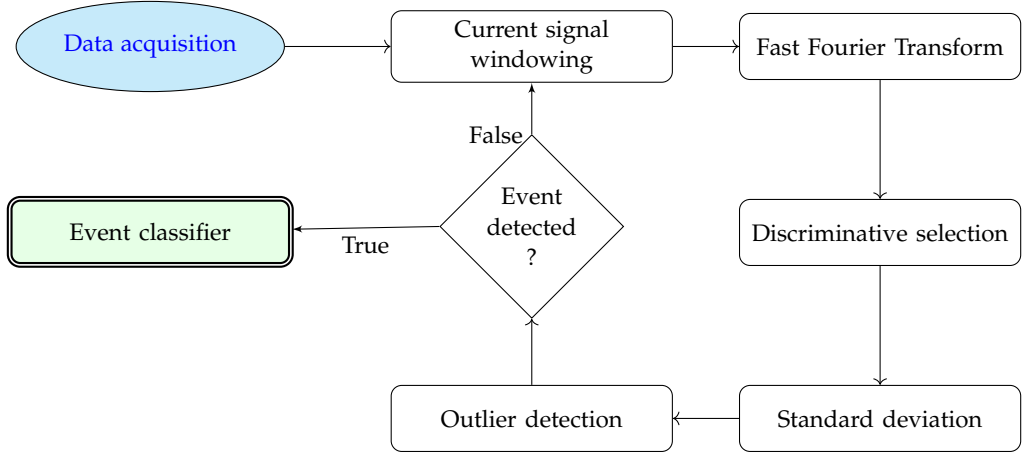
\begin{figure*}[ht!]
	\begin{center}
		\tikzstyle{block2} = [rectangle, draw=black, double, fill=green!10, text width=100 pt, text centered, rounded corners, minimum height=25 pt]
		\tikzstyle{block} = [rectangle, draw=black, text width=100 pt, text centered, rounded corners, minimum height=25 pt]
		\tikzstyle{decision} = [diamond, draw, text width=40 pt, text badly centered, node distance=2cm, inner sep=0pt]
		
		\tikzstyle{line} = [draw, -latex']
		
		\begin{tikzpicture}[node distance = 2cm, auto]
		
		\node[ellipse, draw, text = blue, fill = cyan!20, minimum width = 2cm,minimum height = 1.2cm] (Data) {\small{Data acquisition}};
		\node[below=40 pt, thick, fill=green!10] [block2] (NILM) at (Data.south) {\small{Event classifier}};
		\node[right=40 pt] [block] (Windowing)  at (Data.east) {\small{Current signal windowing}};                    \node[right=25pt] [block] (Fourier)  at (Windowing.east) {\small{Fast Fourier Transform}};
		\node[below=45pt] [block] (Discriminatif)  at (Fourier.south) {\small{Discriminative selection}};
		\node[below=45pt] [block] (Deviation)  at (Discriminatif.south) {\small{Standard deviation}};
		\node[left=25pt] [block] (Outlier)  at (Deviation.west) {\small{Outlier detection}};
		\node[above=25pt] [decision] (decide) at (Outlier.north) {\small{Event detected ?}};
		
		\draw[->] (Data.east) -- (Windowing.west);
		\draw[->] (Windowing.east) -- (Fourier.west);
		\draw[->] (Fourier.south) -- (Discriminatif.north);
		\draw[->] (Discriminatif.south) -- (Deviation.north);
		\draw[->] (Deviation.west) -- (Outlier.east);
		\draw[->] (Outlier.north) -- (decide.south);
		\path [line] (decide.west) -- node {\small{True}}(NILM);
		\path [line] (decide.north) -| node {\small{False}} (Windowing.south);
		\end{tikzpicture}

			\centering
			\caption{Tukey's fences-based event detection }
			\label{fig:fig2}
	
	\end{center}
		\end{figure*}

Upon receiving a power signal as its input, the TFED algorithm is capable of producing a chronological list of events detected. Our solution is fundamentally comprised of five key steps, which are illustrated in Figure  \ref{fig:fig2}. Below, we will delve into each step in detail.
\subsection{Current signal windowing }
To simplify the detection algorithm without sacrificing important information regarding fluctuations in power measurements, the BLUED dataset's sampling rate has been halved from 12 kHz to 6 kHz. The current signal's input is subjected to a sliding window $w(t)$, The technique of sliding windows is often employed in algorithms that deal with data compression, signal processing, and pattern recognition. For instance, in the field of image processing, a sliding window can be utilized to scrutinize smaller parts of an image to recognize distinct features like corners or edges. The sliding window technique is typically implemented by selecting a window size based on the needs of the particular application. The window is then advanced across the data one element at a time, with the data within the window being analyzed and the outcome being preserved or utilized in additional computations. The window is subsequently shifted to the next position, and the procedure is repeated until the whole sequence has been evaluated. When utilizing the sliding window method, it is crucial to take into account the appropriate window size. If the window size is too diminutive, significant characteristics may be disregarded or neglected. Conversely, if the window size is too substantial, the amount of time needed for processing and memory usage will become excessively burdensome.TFED sliding window is specified in Eq.~\ref{eq1}, with a step size of 128 samples. Each window spans 47 blocks, with each block made up of 128 samples, resulting in a total of 6016 samples per window, as stated in Eq.~\ref{mat1},The brief duration of the window plays a crucial role in identifying events that occur closely in time.

\begin{figure}[ht!]
  \centering
  \includegraphics[scale=0.3]{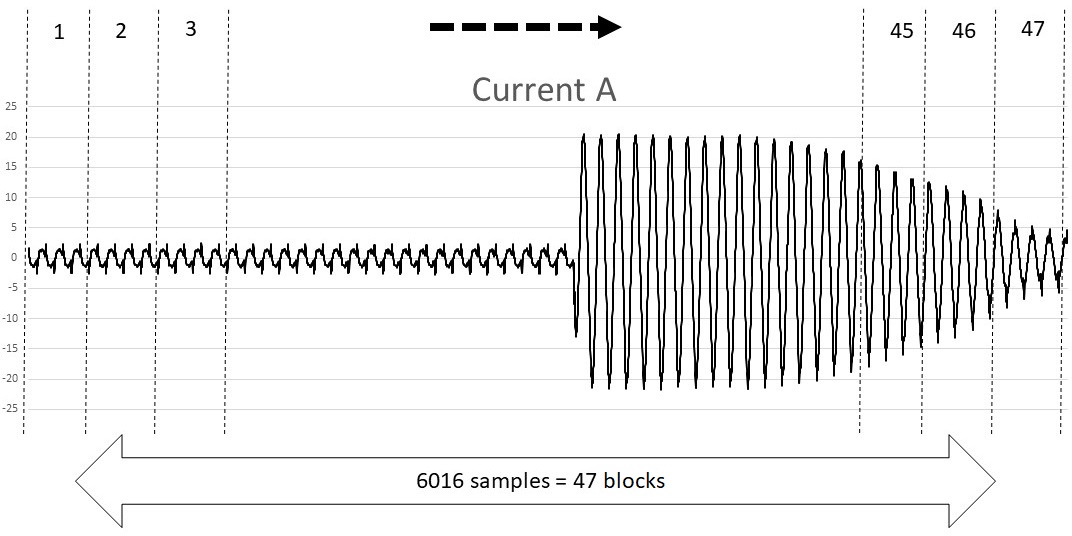}
  \caption{Illustration of a current signal window decomposition.  }
  \label{fig:fig3}
\end{figure}

\newcommand{\Xmatrix}{	\begin{matrix}
			I_{1}  & I_{2}   & \cdots & I_{128} \\
			I_{129} & I_{130} & \cdots & I_{256} \\
			\vdots & \vdots & \ddots & \vdots \\
			I_{4970}& I_{4971}& \cdots & I_{6017}\\
\end{matrix} }

\begin{equation}\label{eq1}
	W(t)= \left( I_{1},I_{2} ,\cdots , I_{6016}\right)
	\end{equation}
	
	\begin{equation}\label{mat1}
	\mathbf{I} = \left( \Xmatrix\right)
	\end{equation}

\subsection{Fast Fourier Transform}
The Fast Fourier Transform, or FFT, is a well-known technique for converting a time-domain signal into a representation of its frequency-domain. This transformation is essential for the analysis of the signal's spectral properties. The FFT processes a finite sequence of samples and outputs a set of complex numbers. Each complex number indicates the amplitude and phase of a specific frequency component within the initial signal.
The FFT is a powerful tool for analyzing signals, as it can examine the frequency composition of a signal to determine important details about its behavior. This includes identifying the dominant frequencies present, detecting the presence of noise or interference, and analyzing the frequency response of a system. Additionally, the FFT can serve as a filter for removing unwanted noise or interference from a signal by eliminating frequency components outside of a specific range. This second step involves performing a fast-forward Fourier transform (FFT) on each block of the current signal that has been windowed. The FFT transforms the signal into spectral components, which are frequency based. It is important to note that the number of current samples should be a power of two. Consequently, the output samples, referred to as frequency bins, are equal to (number of input samples)/2 + 1. This means that with 128 input columns, we will have 67 output samples. All frequencies are expressed in Hertz. Upon completion, the output of the FFT is a row of complex numbers. To simplify further processing, the absolute value of each of these complex numbers is calculated in this step. Figure~\ref{fig:fig3} depicts how this step is executed. The output of this step is represented by matrix $F$ in~\ref{mat2}.
  
	\newcommand{\Ymatrix}{	\begin{matrix}
			f_{1,1} & f_{1,2} & \cdots & f_{1,67} \\
			f_{2,1} & f_{2,2} & \cdots & f_{2,67} \\
			\vdots & \vdots & \ddots &  \vdots   \\
			f_{128,1} &f_{128,2} & \cdots & f_{128,67} \\
	\end{matrix} }
	
    \begin{equation}\label{mat2}
	   \mathbf{F} =\mathbf{FFT} \left( \mathbf{I} \right) = \left( \Ymatrix \right)
	\end{equation}

Each column of the matrix represents a defined frequency. 

\subsection{Discriminative Selection}
After the previous step, a total of 67 frequency domains will be generated. When these frequency domain values are graphed over time, as depicted in Figure 4.5, it becomes apparent that each frequency exhibits its own unique characteristics. The objective of the Discriminative Selection step is to identify the frequency that has the most apparent transition point between the two halves of the window. Once $F$ has been acquired, the calculation of $\Delta P$ for every frequency column is carried out. $\Delta P$ is determined by subtracting the mean of the frequency values on the right side ($\mu_{R}$) from the mean of the frequency values on the left side ($\mu_{L}$), as seen in Equation~\ref{eq2}.
	
\begin{figure}[ht!]
		\centering
		\includegraphics[scale=0.6]{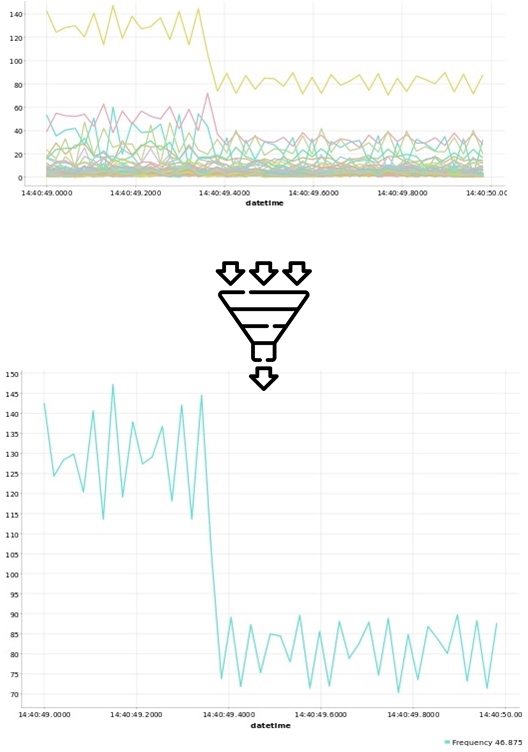}
		\caption{An Illustration of discriminative selection }
		\label{fig:fig4_5}
\end{figure}

\begin{equation}\label{eq2}
	\begin{split}
	\Delta P_{K}&= \left\vert \mu_{L} - \mu_{R}  \right\vert       \quad \textrm{ with} \quad \mu =\frac{1}{N}\sum_{i=1}^{N} X_{i}\\
	&=  \left(\left\vert\frac{1}{64}\sum_{i=1}^{64} f_{i,k} -\frac{1}{64}\sum_{j=65}^{128} f_{j,k} \right\vert \right)      \quad \textrm{for} \quad k  \in \left[1;67\right]
	\end{split}
	\end{equation}
$\mu$ is the arithmetic mean while $N$ is the size of the given set and $X_i$ are the set values
	\begin{equation}\label{eq3}
	\Delta P_{S}= \max \left( \Delta P_{1..67}\right)
	\end{equation}

To select a frequency column, the maximum distance between the means of the first and second halves must be determined. The resulting time series will have 128 samples of the frequency bin, chosen from the 67 blocks outlined in Eq.~\ref{eq3}. The selected column is represented by the index ($S$).

\begin{equation}\label{eq4}
	X_{i} =F_{i,S} \quad \textrm{for} \quad i  \in \left[1;128\right]
\end{equation}
\subsection{Forward Standard Deviation}
To calculate the forward standard deviation, a fixed number of data points are taken and put into a moving window, where the standard deviation is then calculated. This window is then advanced by one data point and the process is repeated for the next set of data points. This is repeated until the entirety of the data series has been analyzed. in this step, we apply a moving forward standard deviation to each frequency bin with a window consisting of four data points, the resulting equation is as follows.  ~\ref{eq6}.
	
	\begin{equation}\label{eq6}
	  \sigma (t) = \sqrt{\frac{1}{4} \sum_{i=t}^{t+3}(x_{i}-\mu)^{2}} \qquad
	\end{equation}

\subsection{Outlier detection}
Outlier detection is a process of identifying observations or data points that deviate significantly from the majority of the data points in a given dataset. These observations are referred to as "outliers" or "anomalies" and can be caused by measurement errors, data processing errors, or rare events. This step aims to identify and address outliers in the input data, which is achieved by utilizing the interquartile range (IQR). Tukey's Fences method is implemented to detect outliers in the input. The first and third quartiles, denoted as $(Q_1, Q_3)$, are computed. Any observation beyond the range R, as represented by the equation ~\ref{eq7}, is deemed an outlier.

	\begin{equation}\label{eq7}
	\begin{split}
	&f(\sigma (t))=
	\begin{cases}
	0,& \text{if }\sigma (t)\in R\\
	1,              & \text{otherwise}
	\end{cases}\\
	\\
	&R=[Q1-k(Q3-Q1)\quad ,\quad Q3+k(Q3-Q1)]\\
	\end{split}
	\end{equation}

We establish the value of $k$ as 0.5, which is typically the minimum value of R that corresponds to the lowermost end of a whisker in a boxplot, while the largest value corresponds to the uppermost end. If an observation is marked as an outlier, the window is also marked as an event window. In the event that all observations are not marked as outliers, the window is marked as a non-event window. It is important to note that any absent values in the data are not taken into consideration and are neither used to calculate outliers nor reported as outliers.
	
	\begin{figure}
		\centering
		\includegraphics[scale=0.45]{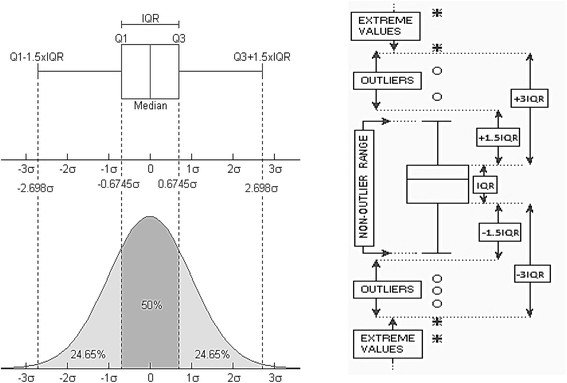}
		\caption{Principle underpinning Tukey's Fences  \cite{tomazs2016}}
		\label{fig:fig6}
	\end{figure}
	
\section{Experimental Results} \label{Simulations}

As far as we are aware, there is no existing literature that assesses the efficacy of event detectors when applied in NILM contexts.
The most effective event detector is undoubtedly the one that produces the most favorable results in terms of disaggregation. If $E_k$ represents the energy consumed by a particular appliance, and $I_k$ represents the estimated energy consumed by the same appliance, then the best algorithm for event detection would be the one that minimizes the difference between these two values, which is demonstrated in Equation~\ref{eq4}.
	
	\begin{equation}\label{eq8}
	\min (\sum_{k=1}^{N} \vert E_{k}-I_{k} \vert)
	\end{equation}
	
Due to the direct correlation between the effectiveness of the event detection and the performance of the NILM classifier, reducing the distance between them is a difficult task. If the event detector is not functioning properly, it will have a negative impact on the implementation of the NILM classifier. Consequently, the most optimal approach is to separate the performance of the event detector from the classification stage to achieve optimal results.
	
In order to test our TFED event detector, we employed the BLUED dataset. During testing, we made use of several evaluation metrics, including TP, FN, FP, Precision, Recall, and F-measure. "Precision" refers to the proportion of detected events that are genuine, while "Recall" denotes the proportion of actual events that are detected.

	\begin{itemize}
		\item  TP are the true-positives (correctly predicted events).
		\item  FP are the false-positives (predicted events that were not real).
		item  FN are the false-negatives (events not detected).
		\item  Precision = TP/(TP+FP).
		\item  Recall = TP/(TP+FN).
		\item  F-measure=(2 $\times$ Precision  $\times$ Recall)/(Precision + Recall).
  \item Accuracy = (TP + TN) / (TP + TN + FP + FN).
	\end{itemize}

\begin{table}[ht!]
		\centering
		\caption{Environment parameters}
		
		\label{tab_1}
		\begin{tabular}{@{}ll@{}}
			\toprule
			Parameter & Value \\ \midrule
			Moving window size & 6016 Samples   \\
			Step size          & 6016 Samples   \\
			K                  & 0.5            \\ \bottomrule
		\end{tabular}
\end{table}

After conducting numerous experiments with various parameters including moving window size, step size, and K constant, best fitting parameters are chosen as demonstrated in Table ~\ref{tab_1}.
One of the primary obstacles encountered in comparing experiments performed on the BLUED dataset in event detection literature is the inconsistency in the number of true events across different papers, as a result of missing values and inaccurate measurement readings. we counted 2444 events in the two phases A and B, the experimental results of the TFED algorithm for event detection. Table \ref{tab_2} presents The results demonstrating that out of the 2444 total events, the algorithm detected 95.3\% true positive events, 0.6\% false negative events, and 1.3\% false positive events. The precision of the algorithm is 99.44\%, which indicates that out of all the events detected by the algorithm, 99.44\% were true events. This is a high precision rate, indicating that the algorithm is effective in detecting true events with minimal false positives.
The recall rate of the algorithm is 98.67\%, indicating that out of all the true events present in the dataset, 98.67\% were detected by the algorithm. This high recall rate indicates that the algorithm effectively detects most of the true events in the dataset, with minimal false negatives.
The F-measure of the algorithm is 99\%, which is a weighted average of the precision and recall rates. This high F-measure score indicates that the algorithm has a high level of accuracy in detecting events.

Comparing our results to other methods, we find that TFED performs favorably. The BLUED method achieves a TPP of 98.2\% but suffers from a higher FPP and FNP compared to TFED. The CUSUM method exhibits lower TPP and higher error rates compared to TFED. The FPGA method demonstrates a lower accuracy and F1 score, indicating potential limitations in its performance.

The GLR method achieves a comparable TPP but significantly higher FPP compared to TFED. References [Kotsilitis2023, Zheng2018] present results that are similar to or slightly lower than TFED in terms of TPP and FPP. WAMMA method achieves a high TPP and accuracy, but the comparison is limited by the lack of available metrics for FPP, FNP, and F1 score, and the number of true events.

Overall, our TFED algorithm performs competitively with existing methods in terms of accuracy and event detection. The results showcase the effectiveness of TFED in accurately identifying events in the BLUED dataset, with low rates of false positives and false negatives. These findings highlight the potential of our proposed approach as a promising solution for event-based NILM systems, contributing to the advancement of load monitoring and energy management applications

\begin{table}[htbp]
\caption{{
Performance comparison with other published work on BLUED.}}
\adjustbox{max width=\columnwidth}{
\begin{tabular}{|l|c|c|c|c|c|c|c|}
\hline
\textbf{Method} & \textbf{TPP(\%) }& \textbf{FPP(\%)} &\textbf{ FNP(\%)} & \textbf{PR(\%)} & \textbf{RE(\%)} & \textbf{F1(\%)} & \textbf{ACC(\%)} \\
\hline
BLUED\cite{anderson2012b} & 98.2 & 2.1 & 1.8 & 98.0 & 98.0 & 98.0 & 96.0 \\
CUSUM\cite{zhu2018a} & 95.8 & 8.1 & 6.6 & 92.2 & 93.6 & 92.7 & - \\
FPGA\cite{Nieto2020} & 85.2 & 1.4 & 14.8 & 98.0 & 85.0 & 91.0 & 84.0 \\
GLR\cite{Luo2002} & 96.7 & 24.0 & 3.3 & 80.1 & 96.7 & 92.7 & - \\
\cite{Kotsilitis2023}  & 94.4 & 2.6 & 5.6 & 97.0 & 94.0 & 96.0 & 92.0 \\
\textbf{TFED} & \textbf{95.3} & \textbf{1.3} & \textbf{0.6 }& \textbf{99.4} & \textbf{98.7} & \textbf{99.0} & \textbf{95.0 }\\
WAMMA\cite{Yan2023} & 99.2 & 0.8 & 0.8 & 99.2 & 99.2 & 99.2 & - \\
\cite{Zheng2018} & 98.7 & 0.7 & 1.3 & 99.0 & 99.0 & 99.0 & 98.0 \\
\hline
\end{tabular}}

\label{tab_2}
\end{table}

Furthermore, TFED surpassed supervised learning approaches that employed Cepstrum analysis and $\chi ^2$ GOF statistic in terms of "F-measure" (Tab.~\ref{tab_3}). However, it must be noted that both supervised approaches exhibited improved performance in processing time, which corresponds to the processing of active power with a low frequency of 60 Hz. We would like to emphasize that a reduction in the step in the current moving window would yield even better results, but would be at the expense of the execution time, which would be considerably extended. In summary, the results demonstrate that TFED is an effective algorithm for event detection, providing high accuracy and demonstrating superior performance compared to other approaches in terms of F-measure.

	\begin{table}[ht!]
		\centering
		\caption{Comparison with other supervised approaches}
		\label{tab_3}
		\begin{tabular}{@{}lc@{}}
			\toprule
			\multicolumn{1}{c}{Algorithm used} & F-measure \\ \midrule
			
			Cepstrum analysis \cite{Norford1996}              & 98\%               \\
			$\chi ^2$ GOF statistic \cite{Norford1996}               & 98\%               \\ 
   TFED (Our Solution)                          & 99\%               \\
   \bottomrule
		\end{tabular}
	\end{table}
	
\section{Conclusion} \label{Conclusion}
To sum up, this academic paper outlines a methodology for detecting events using the TFED algorithm. Results from experiments conducted in this study provide compelling evidence that the TFED algorithm is an extremely effective means for detecting events, with an accuracy rate of 99\%. Additionally, the precision, recall, and F-measure scores all demonstrated high levels of accuracy. Moreover, this approach outperformed other methods such as Cepstrum analysis and $\chi ^2$ GOF statistic supervised learning techniques in terms of F-measure performance.
After conducting a study, it has been suggested that the TFED algorithm may have practical applications in event detection. The algorithm has been proven to possess a high degree of accuracy and dependability in detecting genuine events without producing an abundance of false positives or negatives. This indicates that the algorithm has the potential to be utilized in critical scenarios.
When selecting an algorithm for event detection, it is crucial to achieve a balance between accuracy and speed. The study emphasized this point. Though supervised learning methods showed better processing times, it was the TFED algorithm that demonstrated superior accuracy. As such, the TFED algorithm is the better option for applications where precise event detection is of utmost importance.
The consequences of the research indicate that the TFED algorithm is successful in detecting events, and it has the potential to be applicable in multiple situations where precise event detection is necessary.
After event detection has been completed, the subsequent process involves the extraction of distinct characteristics from transient signals. The integration of these characteristics will result in a more precise identification process in NILM, as well as the creation of genuine and credible energy usage data.
	
 \bibliographystyle{splncs04}
\bibliography{tfed.bib}
	
\end{document}